\documentstyle[11pt,newpasp]{article}

\newcommand{\msun}{\mbox{M$_\odot$}}

\begin{document}

\title{Observations of Circumstellar Disks}

\author{David W. Koerner}
\affil{Dept.\ of Physics and Astronomy, University of Pennsylvania,
Philadelphia, PA 19104-6396}



\begin{abstract}

The imaging of disks around young stars presents extreme challenges
in  high dynamic
range, angular resolution, and sensitivity. Recent instrumental
advances have met these challenges admirably, leading to
a marked increase in imaging discoveries. These have opened
up a new era in studies of the origin of  planetary systems.
Questions about our own solar system's formation, and of the prevalence
of extra-solar planets, are now addressed with  
complementary techniques at
different wavelengths. 
Optical and near-infrared images detail scattered light from 
disks at the highest possible resolution. Mid-infrared, sub-millimeter,
and millimeter-wave techniques probe thermal dust continuum radiation.
Millimeter-wave interferometry details the small-scale structure
of the molecular gas. Kinematic imaging studies affirm the
disk interpretation of mm-wave continuum surveys, and the high
incidence rate for solar nebula analogs. Inner holes, 
azimuthal asymmetries, and gaps suggest the presence
of underlying planetary bodies. The combined techniques provide
a multi-dimensional picture of disks in time and have strengthened
our understanding of the connection between disks and planets.
 Future progress is assured by the
presence of much-improved imaging capability looming on the horizon.

\end{abstract}



\section{Introduction}

Credible observations of disks around young stars 
constitute a rather late entry in the annals  
of important astronomical discoveries, notwithstanding
several earlier misguided attempts. 
Their existence was not firmly established 
until well over two centuries after the
first telescopic observations of galaxies, with which they were occasionally 
confused.  Eighteenth-century speculation that our own solar system may
have formed from a disk is often touted as solely a logical deduction 
from the properties of our own solar system, but it was 
influenced in part by mis-interpretations of 
Herschel's optical observations of galaxies. Laplace, in particular,
 envisioned a {\it hot,}
incandescent nebula that contracted as it cooled, and V.M. Slipher
was still testing this notion in the early twentieth century
when he obtained spectroscopic observations of the Sombrero Galaxy.
The kinematic result conflicted with a nebular interpretation,
but eventually became the velocity half of Hubble's Law
(See Hoyt 1980 for a detailed account). For several decades, the whole
idea of planet formation in a disk fell out of fashion and 
was replaced by  catastrophic theories that discouraged all attempts
to even search for any other example of a forming solar system. 

Today, we know that potentially planet-forming disks are {\it cold} and 
radiate 
predominantly at far infrared wavelengths. Optical imaging can
be used to detect disks only in reflected stellar light and 
at levels that
pale in comparison to the star's direct output (the
ratio of disk/star optical radiation is typically less than 0.0001). 
Such {\it  high contrast} requirements
pose an extreme challenge for optical imaging,
since circumstellar disks at the distance of the nearest star-forming regions
subtend angles of order 1$''$ or less, where glare from the star is
still quite high.
These dynamic range limitations are eased substantially for 
long-wavelength observations
in which disk radiation overpowers that from the stellar photosphere. 
Here, however, {\it angular resolution} becomes a greater challenge, owing to
its inverse proportionality with wavelength for a given aperture. 
High-resolution requirements are
not so severe for a few waning disks around stars that are extremely 
close ($<$ 20 pc), but these tend to radiate very faintly 
and pose a problem in {\it sensitivity}.

In the last two decades, all the obstacles listed above have been met by 
advances in astronomical instrumentation. Coronagraphic 
and Hubble Space Telescope images have surmounted dynamic range challenges;
mm-wave interferometry and mid-infrared detectors at the Keck 10-m
telescope have achieved sub-arcsecond resolution at long wavelengths, 
and bolometer-array imaging at the James Clark Maxwell Telescope has 
provided the requisite long-wavelength sensitivity to image nearby 
disks with low surface brightness.  These techniques have
functioned together as a complementary set of tools for investigating
the properties of
protostellar and protoplanetary disks. The results are reviewed here.
They mark the end of the search for an existence proof for circumstellar
disks, 
but the bare beginning of a {\it new} area of astrophysical research, 
one which promises to be as rich in diverse dynamical examples 
as is the study of galaxies and
perhaps even more important to 
questions about the habitability of the cosmos.
\section{``Indirect'' Observations of Circumstellar Disks}

The first observational evidence for the existence of circumstellar disks 
consisted largely of unresolved detections of long-wavelength emission 
from young stars at levels greater than could be expected from the
stellar photosphere alone.
The measurements were consistent with an origin in circumstellar dust, but 
the geometrical distribution was not apparent. Since a ``disk''  
identification is tantamount to a statement about morphology,
such evidence is considered circumstantial when applied to a disk
interpretation and may be referred to as ``indirect'' 
to distinguish it from imaging results which directly identify a disk-like
shape. This terminology -- direct vs indirect --
refers solely to the evidence for a disk morphology and not to the source
of the radiation. ``Indirect'' observations are routinely interpreted as 
radiation emanating ``directly'' from an unresolved disk. 

Circumstellar dust emission was detected by means of near-infrared 
techniques nearly as soon as these were available (Mendoza 1966). 
Continuum observations taken over a broad spectral range, especially by 
the Infrared Astronomy Satellite (IRAS), were interpreted with models of 
the corresponding Spectral Energy Distribution (SED) to represent snapshots 
in a circumstellar evolutionary sequence (Adams, Lada, \& Shu 1987). 
Infrared point sources with no optical counterpart, often located in 
the center of molecular cloud cores, exhibited SEDs consistent with an 
origin in protostellar radiation reprocessed by both an infall envelope 
and a disk. Excess infrared radiation from 10$^5$ to 10$^7$
year-old T Tauri stars (TTs) matched that expected from viscous 
accretion disks without a surrounding envelope (See reviews by 
Rydgren \& Cohen 1985 and Beckwith \& Sargent 1993); a spherical dust 
configuration was ruled out by comparison of the flux density of
thermal emission with extinction along  the line of sight
to the star (Adams et al.\ 1990;
Beckwith et al.\ 1990). Finally, weak infrared excesses were 
interpreted as tenuous disks in a dispersive stage. 

IRAS detections of infrared excess around nearby main-sequence
stars were initially considered separately from this classification scheme 
(for example, compare review of Backman \& Paresce 1993 with 
Shu et al.\ 1993). However, revised stellar age estimates
 (e.g., Barrado y Navascu\'es et al.\ 1999) and renewed attention to 
the population of 10 million-year-old stars with enhanced debris
disks (e.g., Jura et al.\ 1998) are supporting the idea that many of 
these ``debris disks'' should be associated
with the late stages in the above evolutionary scenario. 

The following sections of this review illustrate that
recent images have largely confirmed the evolving-disk 
scenario implied by unresolved measurements. Indirect observations 
still have an important role to play in disk evolution studies,
however. Surveys of disk properties in statistically significant 
samples are currently available {\it only} with indirect observations. 
These are essential to characterize variations in disk 
properties that depend on something other than time, and that
may mask the identification of evolutionary trends. 
Multiplicity, spectral type, 
cluster environment, and variations in initial disk properties all add 
complications to the task of identifying age-dependent effects. 

Stellar multiplicity has long been considered a factor that may dramatically
affect the formation of planetary systems. Processes of disk formation and 
binary fragmentation both depend on the initial angular momentum budget of
a collapsing cloud core (Bodenheimer 1995). Comparison of 
mm-wave continuum surveys of TTs with the results of speckle interferometric 
binary surveys has given some empirical insight into the nature of the
influence of multiplicity on disks (Mathieu et al.\ 2000).
Dust continuum radiation attributed to disks is reduced for binaries with 
separations of 50-100 AU, similar to a typical disk radius 
(Jensen, Mathieu, \& Fuller 1994; 1996; Osterloh \& Beckwith 1995). 
At both shorter and longer separations, however, the disk detection 
frequency appears to be unchanged. Circumbinary disks around spectroscopic 
binaries (Jensen \& Mathieu 1997), and those around the individual
members of wide binaries (Jensen et al.\ 1996),
appear with the same incidence as for single stars. A few notable exceptions
occur among binaries with $\sim$100 AU separations, including the 
eponymous T Tauri. High-resolution imaging is beginning to
reveal the orientation and likely fate of the material in these systems
(e.g., Akeson, Koerner, \& Jensen 1998; Koerner et al.\ 2000).

Indirect observations of continuum excess from intermediate-mass
stars suggest a disk-evolution scenario similar to that for solar-mass
stars (Hillenbrand 1992). 
The disk interpretation of these measurements has been the subject of
considerable debate, however (see Natta, Grinnin, and Mannings 2000 
for a review). Imaging results indicate that disks around 
Herbig Ae stars (M $<$ 5 \msun) appear to resemble those
around TTs, at least for classical TTs with luminous continuum
excess at millimeter wavelengths (Mannings \& Sargent 1997).
But higher-mass Herbig Be stars  (M $>$ 5 \msun) show little
evidence for similar circumstellar disks, probably as a result
of dispersal under a more energetic radiation environment. 
A rich assemblage of solid-state infrared
features was detected by ISO from the circumstellar environment
of Herbig Ae stars.
Crystalline and amorphous silicates, polycyclic aromatic hydrocarbons,
and unidentified infrared bands differ from typical signatures of
interstellar grains in ways that suggest grain growth and evolution
(see chapter by van Dishoeck, this volume). Sufficient 
sensitivity has not been available to detect similar
spectroscopic features in TTs counterparts. 

Disproportionate sensitivity to properties
of disks around low-mass versus intermediate-mass stars is especially
apparent for infrared excess observations of debris disks. 
IRAS detections of dust around nearby main sequence stars were
largely confined to A stars like Vega, earning them the moniker
``Vega-type'' or ``Vega-excess stars.''
A relative paucity of detections around stars later than
type F should not be construed as evidence of disk absence, however, since 
Vega-type disks would generally fail to radiate above IRAS sensitivity limits 
if placed in the same configuration around all but the very nearest
Sun-like stars. ISO attempts showed a timescale for 400 million years
for A star disks (Habing et al.\ 1999), but efforts to derive similar 
timescales
for Sun-like stars are hampered by small sample size. Nevertheless,
there is some indication that debris disks may survive for {\it longer}
times around later type stars (Song et al.\ 2000). 
This might be explained as the result of
radiation-driven dispersal processes which operate more efficiently around
earlier type stars. Initial attempts to detect the rotational transition
of CO from debris disks suggested that the molecular
gas was dispersed well in advance of this stage
(Dent et al.\ 1995; Zuckerman,
Forveille, \& Kastner 1995). However, these null results may arise
from either photo-dissociation of CO or its depletion onto grains. 
Recent ISO detections of H$_2$ in debris disks 
indicate a normal gas to dust ratio (Thi et al. 2001).

The future of indirect observations is bright. SIRTF will present
an unparalleled opportunity to make spectrosocopic and photometric
measurements of circumstellar dust and gas with unprecedented sensitivity.
From these measurements will arise the first timescales for waning
disks around Sun-like stars. As such, these measurements will greatly
aid our understanding of the connection between disks and planetary
systems! Beyond that, such unresolved measurements will continue to provide
the source lists for imaging efforts, and will yield critical ancillary
information on sources for which imaging has already established the
presence of a disk.

\section{Imaging Disks in Scattered Light}

At optical and near infrared wavelengths, 
high dynamic range constitutes the principal challenge to direct imaging of
circumstellar disks. The stellar luminosity is typically more
than 10$^4$-10$^5$ times
that of the disk  (Whitney \& Hartmann 1992), and  
the wings of the point spread function (PSF)
continue to veil any potential disk emission out to several 
arcseconds from the star. This amounts to several hundred AU for
objects in the nearest star-forming clouds, well beyond the typical
outer disk radius.

Strategies for reducing PSF wing emission include methods which either
occlude the stellar light, re-concentrate it with compensation
for atmospheric blurring, or remove it by subtraction or deconvolution
at the image processing stage. Techniques which have
been applied to disk observations include 1) selection of targets with 
optically-thick edge-on disks which occlude the star, 2) searching
against background nebulosity for disks which would be silhouetted  
(``proplyds''), 3) coronagraphy, 
4) speckle interferometry, 5) adaptive optics, and 6) imaging from space. 
Most of these techniques must be aided by PSF subtraction or deconvolution
with the aid of images of a similar but diskless star. Great care must be 
taken in assuring the uniform conditions for observing the 
target star and PSF. Otherwise, artifacts are 
easily introduced which mimic the appearance of a disk.

Visible and near infrared radiation
is not detected  at the very
earliest stages of circumstellar evolution. It is first  observed when 
optically thick envelope material is sufficiently dispersed
along cavities aligned with the rotational axis. 
A scattering surface reflects the protostellar
radiation as the envelope is flattened about a centrifugally
supported circumstellar
disk. Nebulosity is 
apparent in the surrounding cloud as well and may or may not trace 
the morphology of a circumstellar envelope or disk. In the case
where the flattened disk and envelope are oriented edge-on to the line
of sight, envelope clearing may proceed to an advanced degree before 
light from the central star is directly observed, and the vertical
thickness of the disk may be derived from the silhouette against the
surrounding reflected light. This fortuity 
has been used to good effect with the aid of Hubble Space Telescope
(see articles by Padgett and Stapelfeldt, this volume)
and, in one instance, speckle interferometry at the Keck telescope 
(Koresko 1998). In a similar 
vein, disks have been imaged in silhouette against background
nebulosity such as the Orion Nebula (see review by Beckwith, this
volume).

The detection of light scattered off the surface of a circumstellar 
disk is substantially more difficult if the 
star itself is unobscured by surrounding dust.
High-contrast imaging techniques are essential for success
and have resulted in a handful of  resolved disks around
T Tauri or Herbig Ae stars, including GG Tau (Roddier et al.\ 1996), 
GM Aur (Stapelfeldt et al.\ 1995; Koerner et al.\ 1998), UY Aur
(Close et al.\ 1998), and HD 163296
(Grady et al.\ 1999). Recently, HST WFPC2 (Krist et al.\ 2000)
and NICMOS (Weinberger et al.\ 1999)
observations of a disk around the nearby TTs, TW Hydrae,
were confirmed by Trilling et al.\ (2001) with
ground-based coronagraphic images. The disk is
oriented with its rotational axis parallel to the line of sight.
Displayed in Fig.\ 1, ground-based near-infrared 
coronagraphic images of TW Hya show a face-on disk, in good
agreement with HST/WFPC2 
(Krist et al.\ 1999)
and HST/NICMOS coronagraphic images of the scattered
light (Weinberg et al.\ 1999).
To first order, all three observations produce similar radial 
intensity profiles that
fall off  approximately as the third power of 
the radial distance from the star.

TW Hya
was identified as a TTs by Rucinski \& Krautter (1983). 
It is unusually close by (56 pc), not associated with any molecular cloud, 
and has a surprisingly advanced age (10$^7$ yr) in view of disk properties 
ordinarily associated with much younger stars. The dust 
continuum emission is considerable (Weintraub, Sandell, \& Duncan 1989;
Wilner et al.\ 2000), implying the disk is opaque in the mid-plane with
a total mass like the minimum mass solar nebula ($\sim$0.03 \msun).
Molecular gas is abundant as measured by CO (Zuckerman et al.\ 1994)
and H$_2$ (Weintraub, Kastner, \& Bary 2000). Further, the disk
appears to be still actively accreting onto the star, as inferred
from ample H$\alpha$ emission (Muzerolle et al.\ 2000).
Due to its proximity and 
orientation, TW Hya is a uniquely favorable candidate for studies
of the radial properties
of a viscous accretion disk and will no doubt yield a wealth of 
physical and chemical information
when it is probed by a coming generation of mm-wave arrays that 
can access its low-declination sky position
(e.g., SAO Sub-millimeter Array and NRAO's ALMA).

Ground-based coronagraphic images of the debris disk around
$\beta$ Pictoris marked the first success at direct imaging  
and the first proof of the existence of a plausible
candidate environment
for the formation of planetary systems (Smith \& Terrile 1984).
Initial attempts to duplicate this feat around other IRAS-selected
stars were disappointingly unsuccessful,
leading to speculation that $\beta$ Pic was highly unusual
(Smith, Fountain, \& Terrile 1992; Kalas \& Jewitt 1995).
It now appears, however, that its uniqueness lay mostly
in its combined youth ($\sim$20 Myr; Barrado y Navascu\'es et al.\ 1999)
and proximity (19 pc). Several examples
of similar-age A stars that have comparable fractional excesses are
now known; these indicate 
that $\sim20\%$ of A stars pass through a $\beta$ Pic-like phase 
(Jura et al.\ 1998).  Only $\beta$ Pic is close
enough to be easily imaged with current ground-based coronagraphy,
however (see Fig.\ 9 of Kalas \& Jewitt 1996).

Since its discovery, the disk around $\beta$ Pic has been the
target of diverse detailed imaging studies (see summary in review of 
Lagrange, Backman, \& Artymowicz 2000). The results have 
led to the identification of a number of features
which are hard to understand
without invoking the dynamical influence of one or more substellar
or planet-like
companions. Most telling, perhaps, is a warp that is clearly evident
in HST images (Burrows et al.\ 1995; Heap et al.\ 2000). A distinct
difference in length and brightness
between the two ansae is more difficult to understand,
but is especially apparent in dust continuum images (Lagage \&
Pant\'in 1994; Holland et al.\ 1998). 
Recently, a fine ring structure has been discerned in
modeling of HST images (Kalas et al.\ 2000).

The complementary utility of multi-wavelength
imaging techniques in interpreting scattered-light
images is apparent in images of $\beta$ Pic displayed
in Fig.\ 2.  It is clear, here, that different
wavelengths probe distinctly different
features of the disk. J-band images from ESO, using a 
combined coronagraphic and adaptive optics setup,
show a warp in the disk plane consistent with HST images.
Thermal infrared images at 20 $\mu$m register an asymmetry in
the thickness and length of each of the ansae 
(Koerner et al.\ 2001). Since radiation from
grains with temperatures cooler than 150 K
peak increasingly longward of this wavelength,  the image
registers the distribution of grains with temperatures mostly
above 100~K. Sub-millimeter wavelengths reveal peak emission off
the southwest ansa that is not even apparent at the other
wavelengths. Evidently, massive dust grains at that position are 
not scattering optical light efficiently. It is no surpise, either,
that they are too cold to radiate at thermal infrared
wavelengths. A model of the disk which properly accounts for imaging
at all these wavelengths is sorely needed!

The many features apparent in images of debris disks may testify
of a link between disks and the formation of planetary systems
(see article by Kenyon, this volume).
Persistent dust features which would ordinarily 
disappear on orbital timescales are likely
to be the result of recent dynamical perturbation. If planetary
bodies are the cause, the study of these features is an indirect
method of {\it planet} detection. Just as successful
planet-hunters rely on the motion of
a star to infer the gravitational effect of a planet 
(e.g., Mayor \& Queloz 1995), so it may be that high-resolution
imaging of disks, with the aid of
refined theoretical interpretations, may 
provide clues to the presence of underlying bodies in advance
of their direct detection.

The study of debris disk properties at high resolution received
a recent boost with the installation of the NICMOS camera on
HST. Its coronagraphic capability, coupled with a thorough
program of PSF characterization, has led to imaging of gaps
and holes for the very first time. Key imaging results include 
confirmation of a narrow ring around HR 4796A that was 
originally inferred from model-fitting to thermal infrared images
(Koerner et al.\ 1998; Schneider et al.\ 1999) and
the discovery of a gap in the disk around HD 141569 
(Augereau et al.\ 1999; Weinberger et al.\ 1999). 
In the case of HR 4796A, the images confirmed what was deduced 
from models of the spectral distribution of radiated energy
(Jura et al.\ 1995), namely that the dust surrounded a 
large inner hole. This is readily apparent in thermal IR and HST images
of HR 4796A shown in Figure 3, where the dust is confined largely to a 
circumstellar ring. There is also an asymmetry in the 
brightness distribution evident in HST images. The correct
explanation for these  features is a matter of continued
theoretical investigation (see chapter by Kenyon, this volume).

The recent coronagraphic detection of disks around main sequence
stars with known planets have great potential for further  
strengthening our understanding of
the disk-planet connection. The zodiacal dust
in our own solar system is its most readily detectable feature
from distances of tens of parsecs, and many such disks around Sun-like
stars could have evaded detection by IRAS. The first image of such a
disk was obtained for 55 Cnc by Trilling \& Brown (1998) using
the same instrumentation which imaged TW Hya in Fig.\ 1. 
This was followed by reports
of disks around 3 more such stars (Trilling, Brown, \& Rivkin 1999).
Indirect detections of the disk around 55 Cnc
were first reported in support of ISO observations
by Dominik et al.\ (1998). A recent attempt by 
Schneider et al.\ (2001) to image 55 Cnc
with the HST/NICMOS coronagraph failed 
to confirm the presence of a disk with emission
at the level reported by Trilling \& Brown (1998).
This conflict may be the result of a flux calibration error
in the initial discovery paper (Trilling, private communication).
However, independent confirmation of this  observation remains
highly desirable. Nevertheless, the case for a non-artefactual
origin of disks like those reported by Trilling \& Brown (1998) and Trilling
 et al.\ (1999) is strengthened by the detection of a similar disk by
an independent system. 
The ESO coronagraphic/adaptive-optics system  that produced the
$\beta$ Pic image in Fig.\ 2 (ADONIS) has revealed a disk around
the star $\iota$ Horologii. The image, together with that of a
diskless reference star, is shown in Fig.\ 4.

\section{Thermal Infrared Imaging of Disks}

In the last few years, 
mid-infrared detectors have been incorporated into cameras
suitable for interfacing with large aperture telescopes.
These arrays are capable of detecting thermal
dust radiation at temperatures as low as 100~K.
This technique has
several unique advantages for disk imaging studies. At the corresponding
wavelengths of 10--20 $\mu$m, 
stellar photospheric luminosity is greatly reduced, and high
dynamic range is not as great a challenge. Further, when used
in combination with large aperture telescopes, sub-arcsecond resolution
can be achieved. Although the diffraction
limit is less favorable than for optical wavelengths 
(Airy disk FWHM $\propto \lambda$),  improved seeing
provides considerable compensation at wavelengths longward
of 10 $\mu$m
(Seeing Disk FWHM $\propto \lambda^{-1/5}$). Near 
diffraction-limited resolution
of 0.2--0.4$''$ at $\lambda$ = 10--20 $\mu$m
is routinely achieved using the Keck telescope, for example.
In nearby star-forming clouds, this corresponds to spatial scales of 
30--60 AU. This is generally
insufficient to resolve thermal infrared
emission from accretion disks around Sun-like stars in the T-Tauri phase
or earlier. Effective temperatures of disk surfaces are typically too 
cold at these
radial distances. However, 
the situation is much more favorable for disks
around earlier type stars 
that are not quite so distant. The greater stellar luminosity and
concomitant elevated temperature of the circumstellar dust, coupled with
better spatial resolution due to close proximity, have resulted in 
several exciting new imaging results.

After $\beta$ Pic, the first disk resolved in images by both thermal 
infrared and optical techniques was 
HR 4796A (Koerner et al.\ 1998; Jayawardhana et al.\ 1998;
Telesco et al.\ 2000; Wahhaj et al.\ 2000). The dimensions,
orientation, and morphology of an outer dust ring were first 
inferred in detail with the aid of Bayesian fitting to Keck infrared
imaging (Koerner et al.\ 1998). As evident in Fig.\ 3, 
the result was dramatically confirmed in HST images at higher resolution. 
A brightness asymmetry is clearly evident in the ring at both optical
and infrared wavelengths,
and has been interpreted by Telesco et al.\ (2000) and 
Wyatt et al.\ (1999) as evidence for hidden planets.
This feature is evident in images at 12.5 and 24.5 $\mu$m 
displayed in Fig.\ 5. Other asymmetries are apparent in
independently obtained images as well, including a slight offset between
the ring center and the stellar position. 
In addition, the inner hole
is not completely evacuated of dust; central peak emission in
24.5 $\mu$m image in Fig.\ 5 is in excess over
photospheric emission by factors of several. Rough estimates of
the color temperature locate this dust between 5 and 10 AU from
the star (Wahhaj et al.\ 2000). Much work remains to be done
to securely derive the properties of this system in order
to properly guide theoretical interpretations.

Other disks which have been resolved with mid-infrared techniques
include HD 141569 (Fisher et al.\ 2000; Marsh et al.\ 2001), 
and 49 Cet (Koerner, Guild, \& Sargent 2001). Imaging of HD 141569
forms a great complement to HST images, providing an estimate
of the dust content close to the star where optical techniques
are insensitive. In keeping with its close association with 
temperature, the mid-infrared emission is detected only to a
distance of 100 AU from the star (Fisher et al.\ 2000), in 
contrast to HST imaging which traces material out to a radius of
400 AU (Augereau et al.\ 1999; Weinberger et al.\ 1999).
Keck imaging of dust around 49 Cet traces material out to a
similar distance (Koerner et al.\ 2001); Bayesian modeling of
both the SED and image provide strong evidence for an inner
hole radius of $\sim$20 AU. In contrast to HR 4796A, however,
the SED does not support truncation of the outer radius. 
49 Cet may, in fact, most resemble $\beta$ Pic in its overall
properties. Its greater distance (3 times further away)
and lack of edge-on orientation may be the only reasons
that it has not yet succumbed to coronagraphic imaging attempts.

Of the four disks imaged at thermal infrared wavelengths, all
surround A stars with ages estimated to be about 10 Myr.
This provides strong evidence that $\beta$ Pic is {\it not}
an oddity! But it also points out a detection bias inherent
in infrared techniques to date: they are luminosity limited
at levels which preclude a fair sampling of the dust systems 
around later type stars. 
If disks at this age generally have inner holes with radii of
several times 10 AU, then 
the dust in outer regions is less likely to be heated by
later-type stars to 
levels commensurate with detection at mid-infrared wavelengths.
Fortunately, there are alternatives at even longer wavelengths.

\section{Observations with Submillimeter-wave Telescopes}

Dust continuum radiation from disks typically peaks at far infrared
and sub-millimeter wavelengths. Consequently, imaging at submillimeter 
to millimeter wavelengths is optimum for  
tracing the dust density distribution. 
Single-dish telescopes have very low angular
resolution at these wavelengths, however 
($15''$ at $\lambda$ = 850 $\mu$m for the JCMT). 
Observations of disks in the nearest star-forming regions are 
thus restricted to indirect measurements of continuum flux densities 
(e.g., Beckwith et al. 1990) or molecular-line spectra 
(Dutrey, Guilloteau, \& Guelin 1997). In contrast, the closest debris
disk candidates have expected angular extents of
up to several times the resolution element of single-dish submillimeter
telescopes. Nevertheless, their radiation {\it per unit surface area} is quite
low. The development of highly sensitive bolometer arrays has 
opened a window onto these systems with truly exciting results.

In addition to the image of $\beta$ Pic shown in Fig.\ 2, 
the first images of debris disks around several of the
strongest excess sources (as determined by IRAS) were 
accomplished using the SCUBA bolometer array at the JCMT.
Images of $\alpha$ Psa (Holland et al.\ 1998)
and $\epsilon$ Eri (Greaves et al.\ 1998) provided the
first unambiguous evidence for inner holes in these systms.
SCUBA/JCMT imaging also brought forth the discovery image of a
disk around the first such star identified: Vega.
The presence of circumstellar dust around $\epsilon$ Eri, shown
in Fig.\ 6, is notable on several counts. An inner evacuated
region is clearly evident. And the ring of dust, itself, shows
pronounced azimuthal asymmetries in brightness and thickness.
Theoretical simulations can reproduce these features with the
aid of embedded planetary bodies (Liou \& Zook 1999).
Perhaps most important, however, is the fact that this disk
surrounds a K star with an age of 0.5-1.0 Gyr.
Such a detection is only possible because of the extremely
small distance to the star, 3.2 pc.  
Recent radial velocity measurements strengthen
the disk-planet connection further; a Jupiter-mass planet 
has been detected indirectly at
a radial distance of about 3 AU from the star
(Hatzes et al.\ 2000). If it emerges that
$\epsilon$ Eri is no more unusual than $\beta$ Pic, then the
nearby stellar population is potentially teeming with similar
signatures of planetary systems.

\section{Aperture Synthesis Imaging}

Millimeter and sub-millimeter wavelengths are best
for discerning the properties of  
circumstellar disks at their earliest formation times.
Photospheric emission is entirely negligible, 
dust optical depth is more favorable for probing total mass,
and key molecular species can be identified by  
rotational transitions. Heterodyne measurements
of molecular-line transitions also have the spectral
resolution to enable kinematic studies of gas in the
envelope and disk. At these wavelengths, however,
the requirements on aperture size for the requisite
angular resolution present show-stopping
economic and technical challenges. The most 
successful strategy has been to forsake filled apertures
for interferometric arrays. To date,
these have produced the most
important successes in imaging disks in the gas-rich
accretion phase.

The first image of a gas-rich analog of the
solar nebula followed on the heels of the discovery of a debris 
disk around $\beta$ Pic (Beckwith et al.\ 1986). Aperture synthesis
imaging of HL Tauri carried out with the millimeter
array at Owens Valley Radio Observatory (OVRO) 
revealed a molecular structure with a
diameter of 2000 AU and a velocity structure that
was originally interpreted as Keplerian 
rotation in a disk (Sargent \& Beckwith 1987).
More detailed analyses led to the identification
of an infall component dominating the velocity structure
in the outer regions (Hayashi et al.\ 1993). 
Sub-arcsecond continuum observations were interpreted as tracing
an inner protoplanetary disk with a radius of order 100 AU 
(Lay et al.\ 1994; Mundy et al.\ 1996; Wilner, Ho, \& Rodriguez 1996). 
Flattened structures of similar size and 
kinematics have recently been imaged around other embedded
young stars (see review by Ohashi 2000). As for HL Tau, continuum
observations at the longest wavelengths (3-7 mm) are best at
piercing through envelope material and revealing early disk formation
(see Section IV of review by Mundy, Looney, \& Welch). Sub-arcsecond
continuum imaging of the embedded protostar L1551 IRS5, for
example, has revealed a binary source with individual circumstellar
disks (Looney et al.\ 1997;  Rodriguez et al.\ 1998).

It is at the classical T Tauri phase that aperture synthesis imaging
of disks has been most productive. Kinematic
analysis of OVRO observations of GM Aurigae 
in CO(2$\rightarrow$1) emission
provided the first evidence of solely centrifugal support in a
solar nebula analog (Koerner, Sargent, \& Beckwith 1993). 
Imaging at higher resolution and with improved signal
to noise with the IRAM interferometer confirmed this interpretation
(Dutrey et al.\ 1998). A rapidly growing list of targets have
yielded similar results, including the intermediate-mass star
MWC 48O. CO line emission at distinct velocities is plotted in
Fig.\ 7 together with models of how the emission should appear
if the gas is in Keplerian rotation around the star. The correspondence
provides definitive proof that the gas is confined to a rotating 
circumstellar disk (Mannings, Koerner, \& Sargent 1997). Other
examples for which adequate observations have produced similar results
include GG Tau (Dutrey, Guilloteau, \& Simon 1994),  
DM Tau (Guilloteau \& Dutrey 1998), and LkCa 15 (Koerner \& Sargent 2001).

Imaging surveys mark the continued success of this technique.
Resolved CO emission from DL Tau, RY Tau, DO Tau, and AS 209
also shows velocity gradients in support of a disk interpretation
in OVRO images
(Koerner \& Sargent 1995). Continuum observations of 33 TTs
at $\lambda$ = 2.7 mm were carried out with the IRAM interferometer; 
model fitting to visibilities indicated a typical
angular size 0f 1.5$''$ ($\sim$225 AU) at this wavelength
(Dutrey et al.\ 1996).
An OVRO survey in the CO(2$\rightarrow$1) line was carried
out for the most luminous continuum emitters. Molecular gas emission is
resolved for over 20 classical TTs (Koerner \& Sargent 2001).
These have outer radii in the range 100-600 AU and provide direct
confirmation of the disk interpetation for mm-wave continuum
surveys (e.g., Beckwith et al.\ 1990; Andr\'e \& Montmerle 1994). 
In contrast, an IRAM survey for line emission from weakline TTs
has produced largely null results (Duvert et al.\ 2000). It is
tempting to conclude that this result marks the absence of molecular
gas in these systems, but the recent ISO detection of H$_2$ in
debris disks by Thi et al.\ (2001) suggests this may apply only
to the presence of CO. Indeed, models predict that, as dust settles
to the midplane, stellar photons have sufficient energy to 
photo-dissociate CO. See Dutrey (2000) for a review on depletion of
CO in disks around classical TTs, as well as a discussion of
other molecular species detected in disks.

\section{Future Directions}

High-resolution images of circumstellar disks have
provided insights into the size, morphology, and luminosity of
circumstellar disks in a luminosity-biased sample. Sophisticated
modeling efforts enable a further step to 
estimates of temperature, mass,  chemical composition, and 
dynamical history. In this respect, the field of disk studies
resembles the early history of classification and analysis of
galaxies. However, the most sought-after goal of disk studies
is likely to be our effort to understand the formation mechanisms
and prevalence of planetary systems. In this respect, the most
important future for disk studies lies in making the disk-planet
connection. Many questions remain along a path which should 
eventually see the census of planets, themselves.\par

{\quote { \it

$\bullet$
What is the initial distribution of disk masses and how does it evolve
with time? \par

$\bullet$
What is the timescale for grain growth to planetesimal sizes? \par

$\bullet$
What is the timescale for molecular gas persistence? \par

$\bullet$
What does the gas timescale imply for theories of the 
formation and migration of Jovian planets?\par

$\bullet$
How do stellar and sub-stellar companions affect the dynamic evolution
of disks? \par

$\bullet$
What is the planetary output of typical disks?\par }}
\par
\par
\bigskip

Several instrumental advances lie on the horizon to aid the search
for answers to the above questions. 
Indirect observations will receive an enormous
boost from SIRTF studies aimed at unraveling the dust detection rate
and its early evolution. A large number of team programs, 
including 2 Legacy projects, are poised ready to survey young stars 
and entire molecular clouds for
the presence and character of circumstellar dust emission in
statistically significant samples. From these studies we will find out
the incidence of dust disks around solar-mass stars up to main sequence
ages for the first time. These programs will also provide
useful source lists for a new decade of follow-up imaging
studies at high resolution.

Exciting improvements in high-resolution imaging capability are
looming on the horizon. New adaptive optics systems
and coronagraphic systems are coming on-line together
with a new generation
of large-aperture telescopes.  Optical and infrared interferometric
techniques are also imminent and will
achieve  milli-arcsecond resolution with greatly
enhanced dynamic range. These techniques 
will support the study of disk morphologies
with a view to understanding features diagnostic of undetected
planetary bodies. Aperture sizes of several times 10 m are under
consideration which, when coupled with mid-infrared detectors, will
resolve the emission from TTs and provide a clear picture of AU-scale
details in nearby debris disks. Increased aperture for single-dish
mm-wave studies and improvements in existing mm-wave arrays will
likely provide more than incremental progress in disk studies, since
current studies lie just above a productive detection threshold.
The completion of the Atacam Large Millimeter Array will undoubtedly
launch a heyday in our understanding of the physical and chemical
formation processes at work in the origin of planetary systems.
Taken together, these 
improvements are likely to bring steady fast-paced 
progress in disk studies. As such, they support important scientific
goals which will augment a further revolutionary aim: a thorough
census of planets themselves.

\vfill
\eject

\vskip 1.0truein

\begin{figure}
\vskip 1.0truein

\plotfiddle{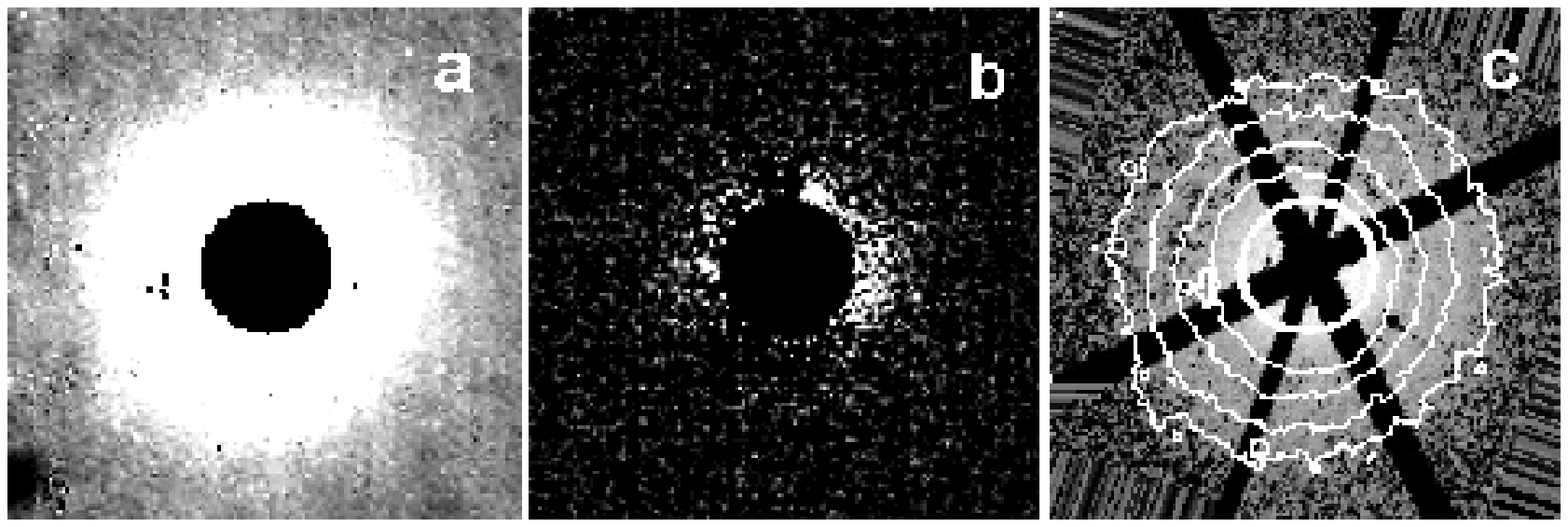}{150pt}{0}{77}{77}{-230}{-215}
\caption {Scattered-light images of the cirumstellar
disk around TW Hydrae from Trilling et al.\ (2001). An H-band 
PSF-subtracted image taken with a ground-based
cooled coronagraph is shown at left. For comparison,
the center panel shows the typical artifact due to
subtraction of two different PSF images. 
The panel at right features the WFPC2 
image of Krist et al.\ (2000) in greyscale
overlain with countours from the ground-based corongraphic
image. All three panels are 7~arcseconds
on a side.}
\end{figure}

\begin{figure} 
\vskip 1.5truein
\plotfiddle{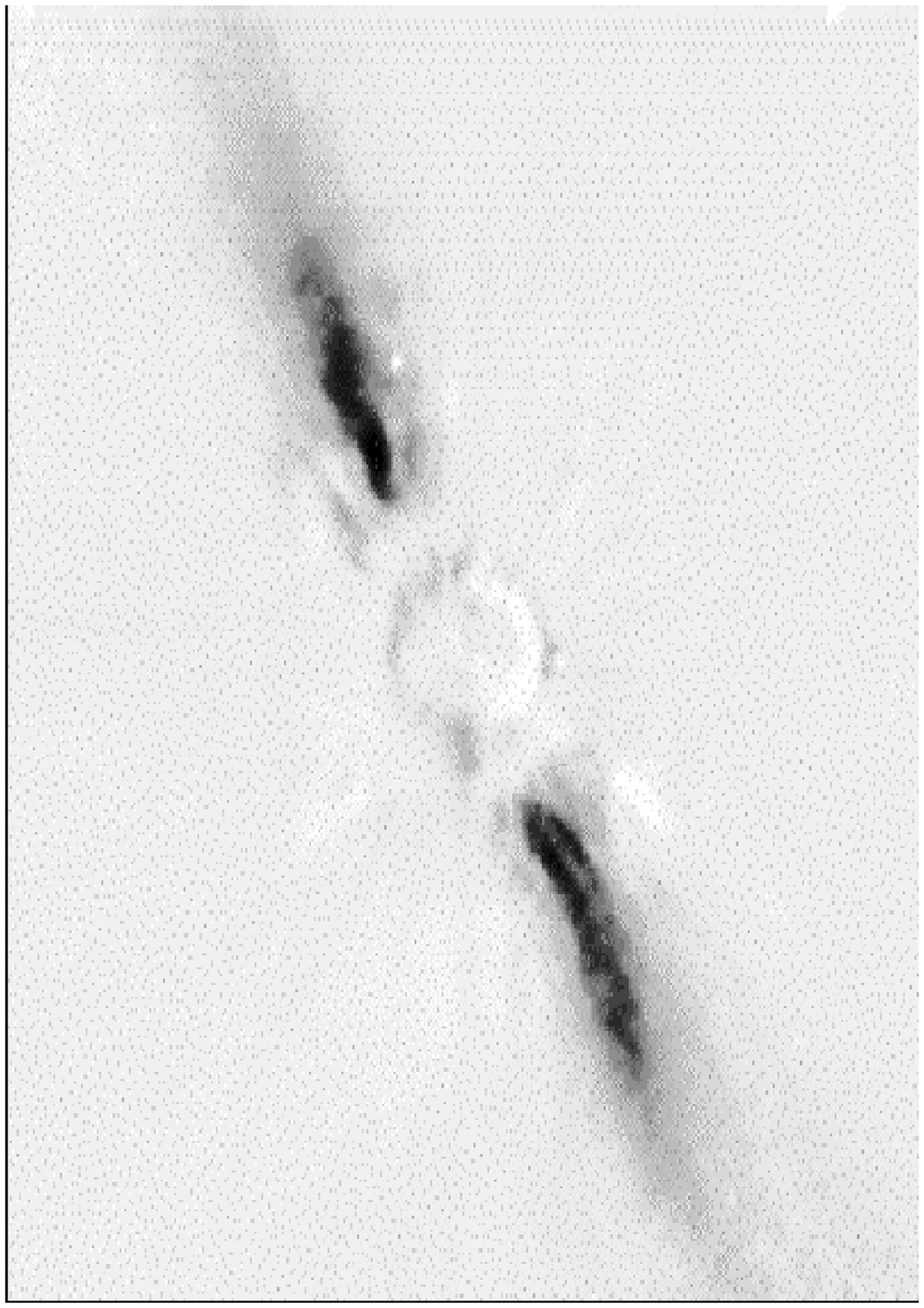}{0pt}{0}{30}{30}{-250}{-100}
\plotfiddle{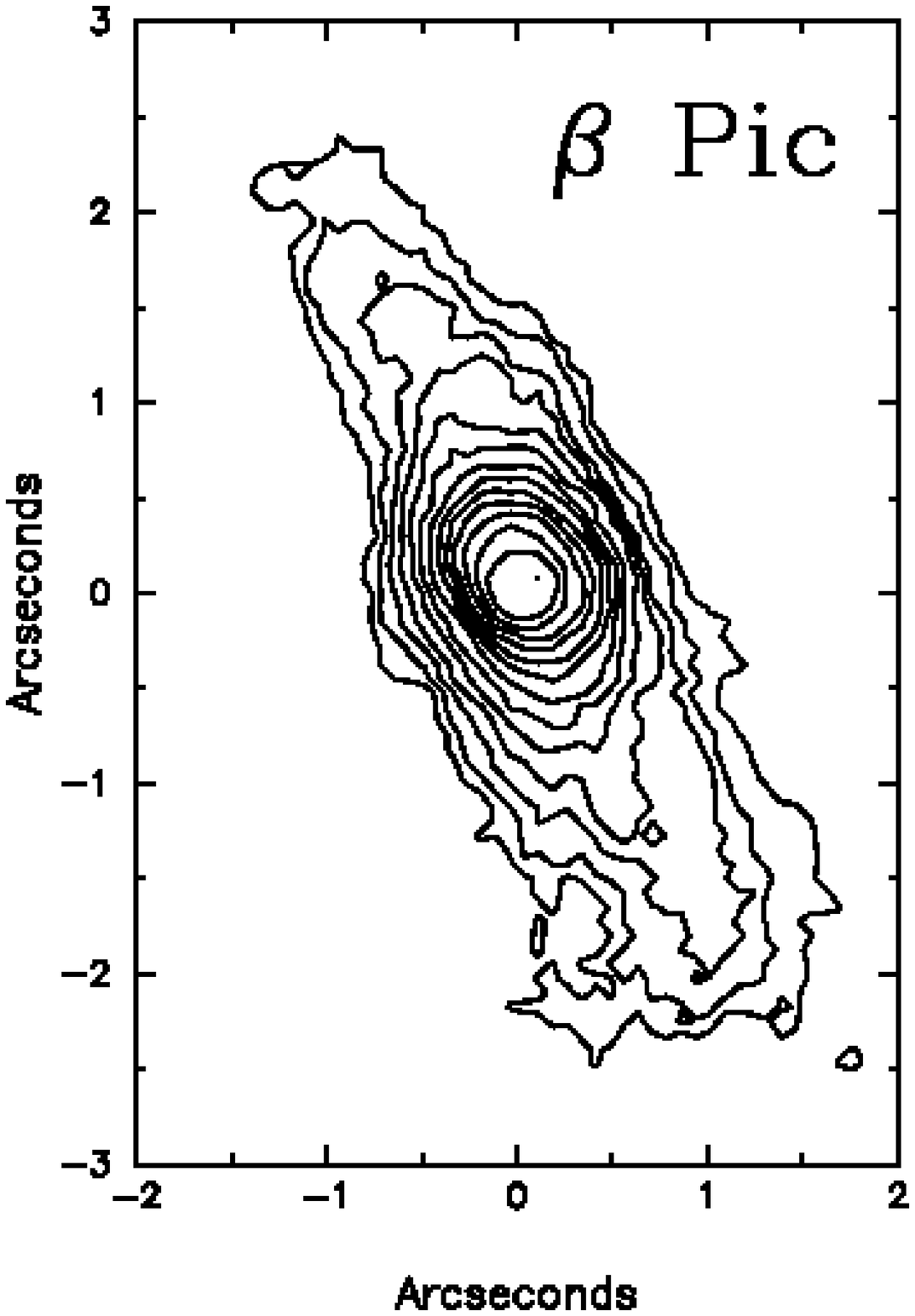}{0pt}{0}{35}{35}{-130}{-100}
\plotfiddle{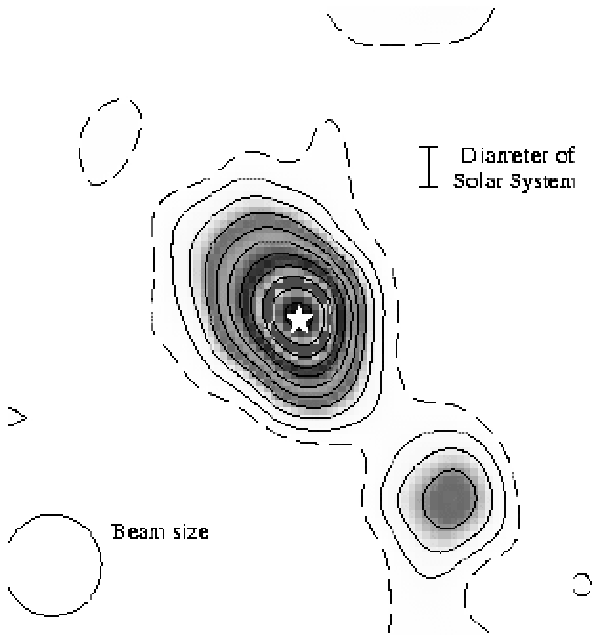}{0pt}{0} {70}{70}{-100}{-200}
\vskip 0.5truein
\caption { Contrasting imaging techniques applied to the
disk around $\beta$ Pictoris.
(Left) J-band  image of scattered light from the circumstellar
disk  acquired by combined coronagraphic and adaptive optics
techniques by ESO. (Center) Mid-infrared image at $\lambda$ = 20 $\mu$m
obtained with JPL's mid-infrared imaging camera, MIRLIN, 
at the Keck telescope. (Right) Sub-millimeter
emission from $\beta$ Pic acquired with the SCUBA bolometer
array at the James Clark Maxwell Telescope.}
\end{figure}

\begin{figure}
\plotfiddle{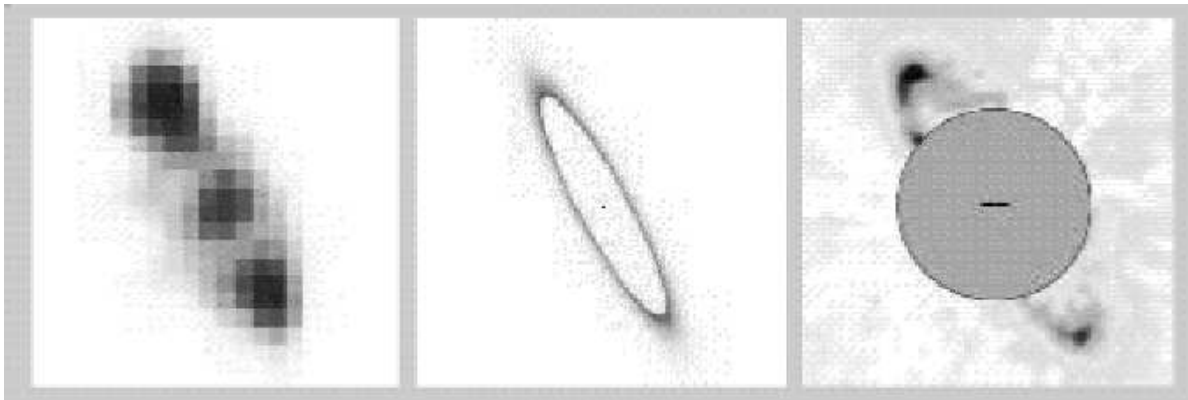}{200pt}{0}{77}{77}{-230}{-190}
\caption {(Left) Keck/MIRLIN image of HR 4796A at 24.5 $\mu$m. The elongated
structure is $\sim$2$''$ in diameter, 
corresponding to $\sim$150~AU. (Center) A model
of the underlying emission structure that was obtained by fitting to
an image at 20 $\mu$m from Koerner et al.\ (1998). 
(Right) HST/NICMOS coronagraphic image of 
scattered light from the ring
around HR 4796A at $\lambda$ = 1.1 $\mu$m taken from Schneider et al.\ (1999).}
\end{figure}
\vskip 2.0 truein

\begin{figure}
\vskip 3.0 truein
\plotfiddle{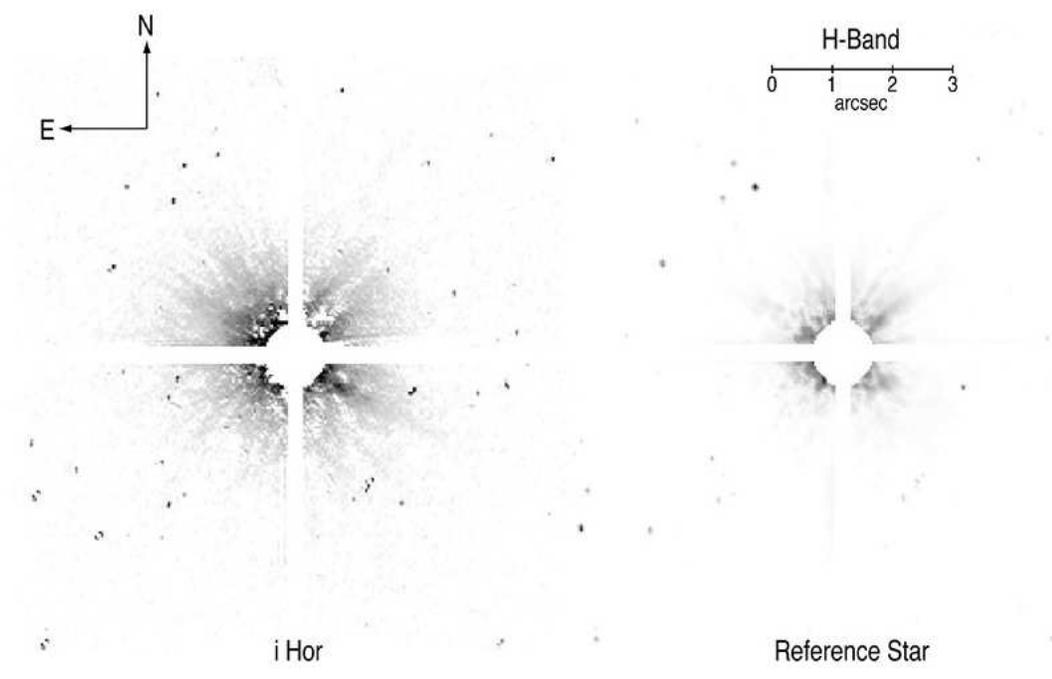}{150pt}{0}{50}{50}{-125}{-50}
\caption{Image of circumstellar dust around the star $\iota$ Horologii (left)
compared to that of a ``reference'' star (right). At a distance of  17 pc, 
$\iota$ Horologii was already known to possess an extrasolar planet. 
Observations were obtained with a coronagraphic mask in conjunction
with the ADONIS adaptive optics  
instrument at the ESO 3.6-m telescope on La Silla.}
\end{figure}

\begin{figure}
\vskip 1.5 truein
\plotfiddle{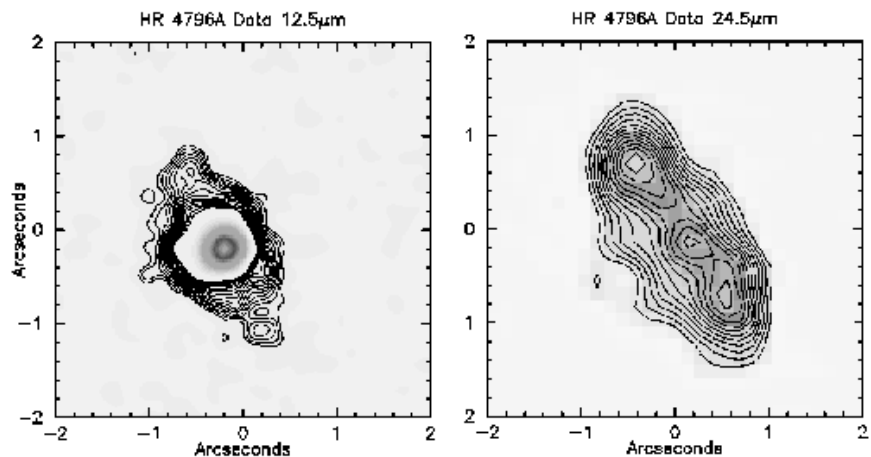}{180pt}{0}{65}{65}{-190}{-150}
\caption {Keck/MIRLIN images of HR 4796A at 12 (left) and 24.5 $\mu$m
(right)
taken from Wahhaj et al.\ 2001. Photospheric emission dominates
the image at 12 $\mu$m, although warm dust associated with the inner
ring can still be detected. At $\lambda$ = 24.5 $\mu$m, the ring
emission has a larger radius, commensurate with that of HST images,
and emission is detected at the stellar position at levels 
substantially above the expected photospheric contribution.
}

\end{figure}

\begin{figure}
\plotfiddle{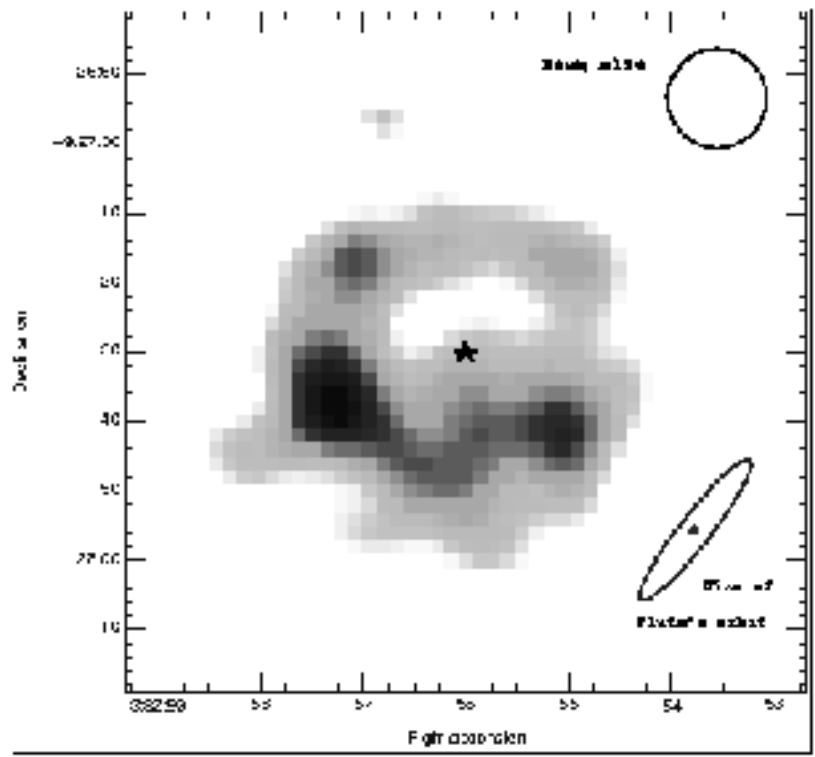}{250pt}{0}{100}{100}{-300}{-190}
\caption {Taken from Greaves et al.\ (1998), a
SCUBA/JCMT 850 $\mu$m image of dust around $\epsilon$ Eri, a
nearby K star for which radial velocity techniques have
identified a Jupiter-mass planet at a distance of 3.2 AU from 
the central star.
}
\end{figure}

\begin{figure}
\plotfiddle{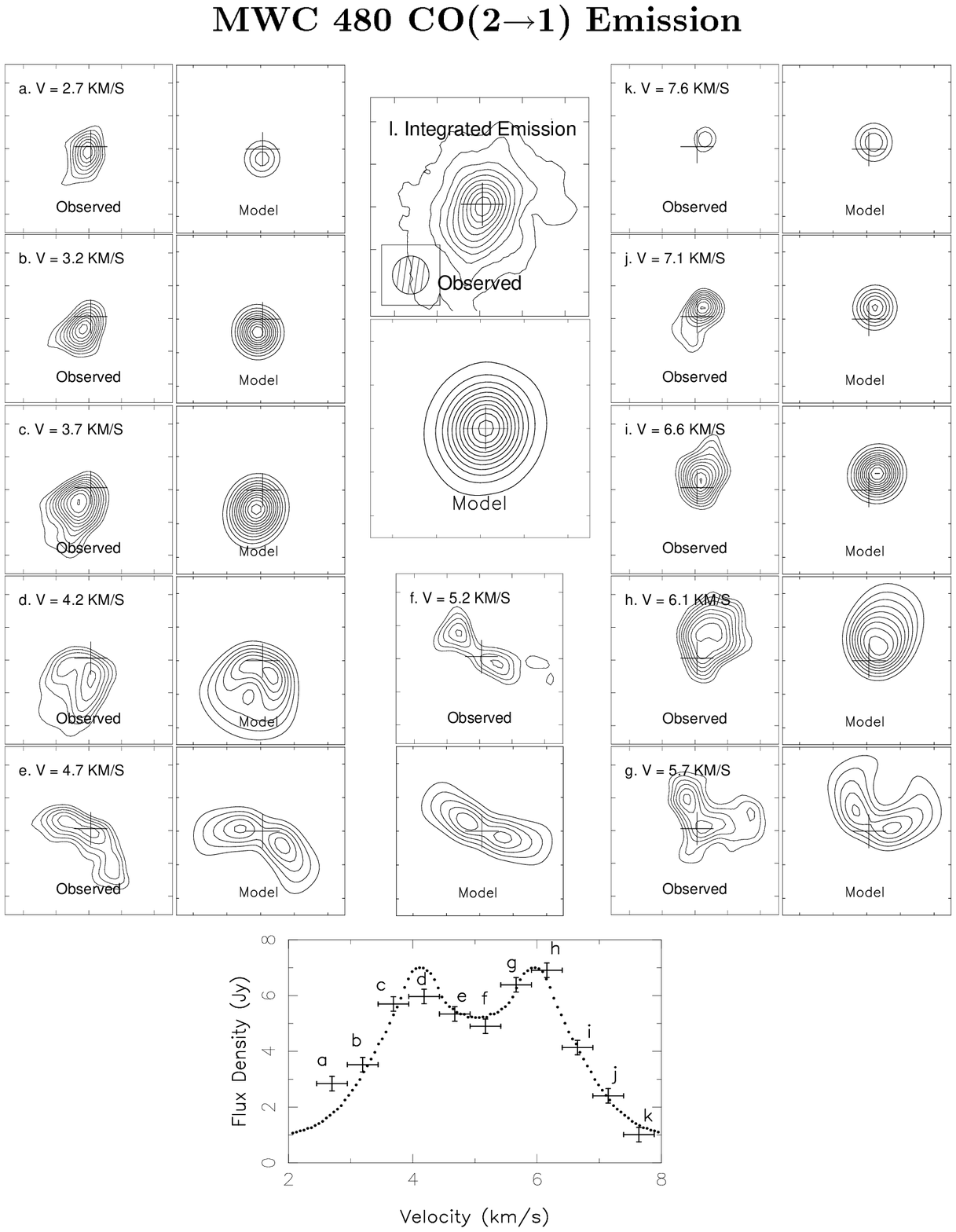}{500pt}{0}{85}{85}{-245}{-75}
\caption {Taken from Mannings, Koerner, \& Sargent (1997),
spectral line maps of CO(2$\rightarrow$1) emission are shown
adjacent to simulations of the emission predicted by a
kinematic model of a disk in Keplerian rotation. A contour plot of
the velocity-integrated emission is displayed in the upper
central panel above the total emission from the model. The resulting
model spectrum is plotted at the bottom,
together with the total flux from each
of the maps.
}
\end{figure}

%


\begin{references}
\reference Adams, F. C., Lada, C. J., \& Shu, F. H. 1987, \apj, 312, 788
\reference Adams, F.C., Emerson, J.P. and Fuller, G.A., 1990, \apj, 357, 606
\reference Akeson, R.,L., Koerner, D.W., \& Jensen, E.L.N., 1998, \apj, 505, 
358
\reference Andr\'e, P., and Montmerle, T.,  1994, \apj,  420, 837
\reference Augereau, J.C., Lagrange, A.M., Mouillet, D., \& M\'enard, F.,
1999, A\&A, 350, L51
\reference Backman, D.E. Witteborn, F.C.., \& Gillett, F.C., 1992, 
\apj, 385, 670
\reference Backman, D.E. \& Paresce, F. 1993, in {\it Protostars and 
Planets III,} eds.\ E.H. Levy and J.I. Lunine,  (Tucson: University 
of Arizona Press), 1253
\reference Barrado Y Navascu\'es, D., Stauffer, J.R., Song, I., \& Caillault, 
J.-P., 1999, \apj, 520, L123
\reference Beckwith, S.V.W., \& Sargent, A.I. 1996, Nature, 383, 139
\reference Beckwith, S. V. W., \& Sargent, A. I. 1993, in {\it Protostars and
Planets III}, eds.\ E. H. Levy \& J. I. Lunine
(Tucson:University of Arizona Press), 521
\reference Beckwith, S.V.W., Sargent, A. I.,  Scoville, N.Z.,
Masson, C.R., Zuckerman, B., \& Phillips, T.G. 1986, \apj, 309, 755
\reference Beckwith, S.V.W., Sargent, A. I., Chini, R., \& G\"usten, R.
  1990, AJ, 99, 924
\reference Bodenheimer, P,: 1995, \araa, 33, 199
\reference Burrows, C.J., Krist, J.E., Stapelfeldt, K.R., \& WFPC2 IDT
1995, BAAS, 187, 32.05
\reference CLose, L.M., Dutrey, A., Roddier, F., Guilloteau, S., Roddier,
C., Northcott, M., Menard, F., Duvert, G., Graves, J.E., \& Potter, D., 1998,
\apj, 499, 883
\reference Dent, W.R.F., Greaves, J.S., Mannings, V.,
Coulson, I.M., \& Walther, D.M., 1995, MNRAS, 277, L25
\reference Dutrey, A., Guilloteau, \& Simon, M., 1994, A\&A, 286, 149
\reference Dutrey, A., Guilloteau, S., Duvert, G., Prato, L., Simon, M.,
Schuster, K., \& M\'enard, F., 1996, A\&A, 309, 493
\reference Dutrey, A., Guilloteau, S., \& Gu\'elin, M., 1997, \aa, 317, L55
\reference Dutrey, A., Guilloteau, S., Prato, L., Simon, M., Duvert, G.,
Schuster, K., \& M\'enard, F. 1998, A\&A, 338, L63 
\reference Dutrey, A., 2000, in IAU Symp. 197, {\it Astrochemistry:
From Molecular Clouds to Planetary Systems},  (San Francisco: PASP)
ed. Y.C., Minh \& E.F. van Dishoeck, 415
\reference Duvert, G., Guilloteau, S., M\'enard, F., Simon, M., \&
Dutrey, A., 2000, \aa, 355, 165
\reference Fisher, R.S., Telesco, C.M., Pi\~na, R.K., Knacke, R.F.,
\& Wyatt, M.C., 2000, \apj, 532, L141
\reference Grady, C.A., Devine, D., Woodgate, B., Kimble, R., Bruhweiler, F.C.,
Boggess, A., Linsky, J.L., Plait, P., Clampin, M., \& Kalas, P., 1999, BAAS,
195, 2.08
\reference Greaves, J.S., Holland, W.S., Moriarty-Schieven, G., Jenness, T.,
Dent, W.R.F., Zuckerman, B., McCarthy, C., Webb, R.A., Butner, H.M.,
Gear, W.K., \& Walker, H.J., 1998, \apj, 506, L133
\reference Guilloteau, S., \& Dutrey, A., 1998, A\&A, 339, 467
\reference Habing, H.J., Dominik, C., Jourdain de Muizon, M., Kessler, M.F.,
Laureijs, R.J., Leech, K., Metcalfe, L., Salama, A., Siebenmorgen, R., 
\& Trams, N.,  1999, Nature, 401, 456
\reference Hatzes, A.P., Cochran, W., McArthur, B., Baliunas, S., Walker, G.,
Campbell, B., Irwin, A., Yang, S., Kuerster, M., Endl, M., Els, S., 
Butler, P., \& Marcy, G., 2000, \apj, in press
\reference Hayashi, M., Ohashi, N., \& Miyama, S.M., \apj, 418, L71
\reference Heap, S.R., Lindler, D.J., Woodgate, B., \& STIS ID Team, 1997, 
BAAS, 191, 47.02
\reference Hillenbrand, L.A., Strom, S.E., Vrba, F.J., \& Keene, J., 1992,
\apj, 397, 613
\reference  Holland, W.S., Greaves, J.S.,
Zuckerman, B., Webb, R.A., McCarthy, C., Coulson, I.M., 
Walther, D.M., Dent, W.R.F., Gear, W.K., \& Robson, I., 1998, Nature,
392, 788
\reference Hoyt, W.G., 1980, {\it Planets X and Pluto}, 
(Tucson: University of Arizona Press)
\reference Jayawardhana, R., Fisher, S., Hartmann, L., Telesco, C., 
Pina, R., \& Fazio, G., 1998, \apj, 503, L79
\reference Jensen, E.L.N., Mathieu, R.D., \& Fuller, G.A. 1994, \apj, 429, L29
\reference Jensen, E.L.N., Mathieu, R.D., \& Fuller, G.A. 1996, \apj, 458, 312
\reference Jensen, E.L.N.,\&  Mathieu, R.D., 1997, \aj, 114, 301
\reference Jura, M., Ghez, A.M., White, R.J., McCarthy, D.W., Smith, R.C.,
\& Martin, P.G., 1995, \apj, 445, 451
\reference Jura, M., Malkan, M., White, R., Telesco, C., Pina, R., \& 
Fisher, R.S., 1998, \apj, 505, 897
\reference Kalas, P., \& Jewitt, D., 1995, Ap\&SS, 223, 167
\reference Kalas, P., \& Jewitt, D., 1996, \aj, 111, 1347
\reference Kalas, P., Larwood, J., Smith, B.A., \& Schultz, A., 2000, \apj, 530, L133
\reference Koerner, D.W. 1997, in {\it Planetary and Interstellar Processes
Relevant to the Origins of Life}, ed. D.C.B. Whittet, (Kluwer: Dordrecht)
pp. 157
\reference Koerner, D.W. \& Sargent, A.I. 1995, AJ, 109, 2138
\reference Koerner, D.W., Sargent, A.I., \& Beckwith, S.V.W. 1993, Icarus, 
106(1), 2 
\reference Koerner, D.W., Ressler, M.E., Werner, M.W., \& Backman, D.E. 1998, 
ApJ, 503, L83
\reference Koerner, D.W., Schneider, G., Smith, B. A., Becklin, E. E., 
Hines, D. C., Kirkpatrick, J. D., Lowrance, P. J., Meier, R., Reike, M., 
Terrile, R. J., Thompson, R. I., 1999, BAAS, 193, 73.14
\reference Koerner, D.W., Jensen, E.L.N., Cruz, K.L., Guild, T.B., \&
Gultekin, K., 2000, \apj, 533, L37
\reference Koerner, D.W., Guild, T.B., \& Sargent, A.I., 2001, \apj, submitted
\reference Koerner, D.W., Backman, D.E., Werner, M.W., \&
Ressler, M.E., 2001, in preparation
\reference Koerner, D.W., \& Sargent, A.I., 2001, in preparation
\reference Krist, J.E., Stapelfeldt, K.R., M\'enard, F., Padgett, D.L., \& 
Burrows, C.J., 2000, \apj, 538, 793
\reference Lagage, P.O., \& Pantin, E., 1994, Nature, 369, 628
\reference Lagrange, A-M., Backman, D.E., \& Artymowicz, P., 2000, 
in {\it Protostars and Planets IV,} eds.\ V. Mannings, A.P. Boss, \& S.S.
Russell (Tucson: University of Arizona Press), 639
\reference Lay, O.P., Carlstrom, J.C., Hills, R.E., \& Phillips, T.G., 1994, 
ApJ, 434, L75
\reference Liou, J.-C., \& Zook, H.A., 1999, \aj, 118, 580
\reference Lunine, D.W. 1997, in {\it Planetary and Interstellar Processes
Relevant to the Origins of Life}, ed. D.C.B. Whittet, (Kluwer: Dordrecht)
pp. 205
\reference Lynden-Bell, D., \& Wood, R.  1968, \mnras, 138, 495
\reference Mannings, V., \& Sargent, A.I. 1997, \apj, 490, 792
555
\reference Mannings, V., Koerner, D.W., \& Sargent, A.I. 1997, Nature, 388,
555
\reference Mathieu, R.D., Ghez, A.M., Jensen, E.L.N., \& Simon, M., 2000,
in {\it Protostars and Planets IV,} eds.\ V. Mannings, A.P. Boss, \& S.S.
Russell (Tucson: University of Arizona Press), 703
\reference Marsh, K.A., Silverstone, M.D., Becklin, E.E., Koerner, D.W.,
Werner, M.W., \& Ressler, M.E., 2001, \apj, submitted
\reference Mayor, M., \& Queloz, D., 1995, Nature, 348, 355
\reference Mendoza, V.E.E., 1966, \apj 143, 1010
\reference Mundy, L.G., Looney, L. W., Erickson, W., Grossman, A., 
Welch, W.J., Forster, J. R., Wright, M.C.H., Plambeck, R.L., 
Lugten, J., \& Thornton, D.D., 1996, ApJ, 464, L169
\reference Mundy, L.G., Looney, L.W., \& Welch, W.J., 2000, 
in {\it Protostars and Planets IV,} eds.\ V. Mannings, A.P. Boss, \& S.S.
Russell (Tucson: University of Arizona Press), 355
\reference Muzerolle, J., Calvet, N., Brice\~no, C., Hartmann, L.,
\& Hillenbrand, L., 2000, \apj, 535, L47
\reference Myers, P.C., \& Benson, P.J., \apj, 248, 87
\reference Natta, A., Grinin, V., \& Mannings, V., 2000, 
in {\it Protostars and Planets IV,} eds.\ V. Mannings, A.P. Boss, \& S.S.
Russell (Tucson: University of Arizona Press), 559
\reference Ohashi, N., 2000, in IAU Symp. 197, {\it Astrochemistry:
From Molecular Clouds to Planetary Systems},  (San Francisco: PASP)
ed. Y.C., Minh \& E.F. van Dishoeck, 61
\reference Osterloh, M., \& Beckwith, S. 1995, ApJ, 439, 288
\reference Padgett, D.L., Brandner, W., Stapelfeldt, K.R., Strom, S.E.,
Terebey, S., Koerner, D. 1999, AJ, 117, 1490
\reference Pantin, E., Lagage, P.O., \& Artymowicz, P., 1997, A\&A, 327, 1123
\reference Roddier, C., Roddier, F., Northcott, M.J., Graves, J.E.,
\& Jim, K., 1996, \apj, 463, 326
\reference Rodriguez, L.F., D'Alessio, P., Wilner, D.J., Ho, P.T.P.,
Torrelles, J.M., Curiel, S., Gomez, Y., Lizano, S., Pedlar, A., Canto, 
J., \& Raga, A.C., 1998, Nature, 395, 355
\reference Rucinski, S.M., \& Krautter, J., 1983, RMxAA, 7, 200
\reference Rydgren, A.E., \& Cohen, M., in  
{\it Protostars \& Planets III}, ed. E. H. Levy \& J. I. Lunine 
(Tucson: University of Arizona Press), 371
\reference Sargent, A.I., \& Beckwith, S. 1987, ApJ, 323, 294
\reference Schneider, G., Smith, B.A., Becklin, E.E., Koerner, D.W.,
Meier, R., Hines, D.C., Lowrance, P.J., Terrile, R.J., Thompson, R.I.,
\& Rieke, M. 1999, \apj, 513, L127
\reference Schneider, G.,  Becklin, E., Smith, B.,Weinberger,
A., Silverstone, M., \& Hines, D., 2001, \aj, in press
\reference Shu, F., Najita, J., Galli, D., and Ostriker, E. 1993, in  
{\it Protostars \& Planets III}, ed. E. H. Levy \& J. I. Lunine 
(Tucson: University of Arizona Press), 3
\reference Smith, B.A., \& Terrile, R.J. 1984, Science, 226, 1421
\reference Smith, B.A., Fountain, J.W., \& Terrile, R.J., 1992, \aa, 261, 499
\reference Song, I., Caillault, J.-P., Barrado y Navascu\'es, D.,
Stauffer, J.R., \& Randich, S., 2000, \apj, 533, L41
\reference Stapelfeldt, K., Burrows, C.J., Koerner, D., Krist, J., WFPC2 IDT, 
1995, BAAS, 191, 92.04
\reference Telesco, C.,M., Fisher, R.S., Pi\~na, R.K., Knacke, R.F., 
Dermott, S.F., Wyatt, M.C., Grogan, K., Holmes, E.K., Ghez, A., M.,
Prato, L., Hartmann, L.W., \& Jayawardhana, R., 2000, \apj, 530, 329 
\reference Thi, W.F., Blake, G.A., van Dishoeck, E.F., van Zadelhoff, G.J.,
Horn, J., Becklin, E.E., Mannings, V., Sargent, A.I., van Dan Ancker, M.E.,
\& Natta, A., 2001, Nature, in press
\reference Trilling, D.E., \& Brown, R.H. 1998, Nature, 395, 775
\reference Trilling, D.E., Brown, R.H., \& Rivkin, A.S., 1999, \apj, 529, 499
\reference Trilling, D.E., Koerner, D.W., Barnes, J., 
Ftaclas, C., \& Brown, R. H. 2001, \apj,  in press
\reference Wahhaj, Z., Koerner, D.W., Backman, D.E., Werner, M.W.,
Serabyn, E.,\& Ressler, M.E. 2000, BAAS, 195, 25.04
\reference Weinberger, A.J., Schneider, G., Becklin, E.E., Smith, B.A.,
\& Hines, D.C., 1999, BAAS, 194, 69.04
\reference Weinberger, A.J., Becklin, E.E., Schneider, G., Smith, B.A.,
Lowrance, P.J., Silverstone, M.D., Zuckerman, B., \& Terrile, R.J., 1999.
\apj, 525, L53
\reference Weintraub, D.A., Sandell, G., and Duncan, W.D.: 1989, 
\apj, 340, L69
\reference Weintraub, D.A., Kastner, J.H., \& Bary, J.S., 2000, 
\apj, 541, 767
\reference Whitney, B.A. and Hartmann, L., 1992, \apj, 395, 529
\reference Wilner, D.J., Ho, P.T.P., Kastner, J.H., \& Rodr\'iguez, L.F, 2000,
534, L101
\reference Wyatt, M.C., Dermott, S.F., Telesco, C.M., Fisher, R.S.,
Grogan, K., Holmes, E.K., \& Pi\~na, R.K., 1999, \apj, 527, 918
\reference Zuckerman, B., Forveille, T., \& Kastner, J.H., 1995, Nature,
373, 494
\end{references}
\end{document}